\documentclass[prb,twocolumn,showpacs]{revtex4}
\usepackage{graphicx}
\usepackage{amsmath}
\usepackage{amssymb}
\usepackage{dcolumn}
\usepackage{float}
\usepackage{bm}

\DeclareMathAlphabet{\bi}{OML}{cmm}{b}{it}

\def\be{\begin{equation}}
\def\ee{\end{equation}}
\def\bearr{\begin{eqnarray}}
\def\eearr{\end{eqnarray}}

\def\bs{\boldsymbol}

\begin{document}

\title{Phonon-drag magnetothermopower in Rashba spin-split 
two-dimensional electron systems}

\bigskip

\author{Tutul Biswas and Tarun Kanti Ghosh}
\normalsize
\affiliation
{Department of Physics, Indian Institute of Technology-Kanpur,
Kanpur-208 016, India}
\date{\today}
 
\begin{abstract}
We study phonon-drag contribution to the thermoelectric power 
in a quasi-two-dimensional electron system confined in GaAs/AlGaAs 
heterostructure in presence of both Rashba spin-orbit interaction and
perpendicular magnetic field at very low temperature. 
It is observed that the peaks in the phonon-drag thermopower split 
into two when the Rashba spin-orbit coupling constant is strong.
This splitting is a direct consequence of the Rashba spin-orbit interaction.
We show the dependence of phonon-drag thermopower on both magnetic field and 
temperature numerically.
A power-law dependence of phonon-drag magnetothermopower on the temperature 
in the Bloch-Gruneisen regime is found. We also extract the exponent of 
the temperature dependence of phonon-drag thermopower for different 
parameters like electron density, magnetic field, and the spin-orbit coupling
constant.

\end{abstract}


\maketitle

\section{Introduction}
Low-temperature measurement of thermoelectric power provides 
an important tool for probing electronic and transport 
properties of various low-dimensional systems. Extensive investigations 
on magnetothermopower measurement\cite{fletcher1, davidson, obloh, vuong1,
vuong2, fletcher2, fletcher3, lyo1, fletcher4}
in two-dimensional electron system (2DES),
formed at the interface of GaAs/AlGaAs heterostructure, were initiated in 
mid-eighties. Electron diffusion and phonon-drag 
are the two additive contributions to the thermopower. The drift motion of 
electrons, due to external perturbations like temperature gradient or electric 
field, is entirely responsible for the diffusion thermopower. On the other hand 
phonon-drag thermopower originates as an outcome of the interaction between 
electrons and phonons. In low density semiconducting systems a tiny fraction 
of acoustic phonons with wave vector $q \leq 2k_F$ (where $k_F$ is the Fermi 
wave vector) interact with electrons below a certain characteristic temperature
$T_{BG}=2\hbar v_sk_F/k_B$ (where $v_s$ is the sound velocity) because of the 
phase space restriction. The temperature regime defined by $T\leq T_{BG}$ is 
usually known as the Bloch-Gruneisen\cite{stormer,dassarma} (BG) regime. 
In BG regime the diffusion thermopower $S_d$ varies linearly with temperature
whereas the phonon-drag thermopower $S_g$ shows a power-law dependence 
$S_g\sim T^{\delta_e}$, where the effective exponent of the temperature 
dependence $\delta_e$ varies for different systems as well as for different
scattering mechanisms of electron-phonon interaction. 
Past experimental\cite{ruf} and theoretical\cite{sankeshwar} works established 
that $S_d(S_g)$ dominates over $S_g(S_d)$ at temperature below(above) $1$ K. 
In inversion asymmetric semiconducting heterostructures two different 
mechanisms are responsible for the electron-phonon interaction. 
They are known as deformation potential (DP) and piezoelectric (PE) scattering 
potential. Lattice deformation leads to the potential energy change of electrons 
to produce DP scattering potential. On the other hand potential energy corresponding 
to the induced electric polarization due to crystal vibrations is known as PE 
scattering potential.

There are two equivalent methods available in the literature for the calculation
of the phonon-drag contribution to the thermoelectric power. 
According to Herring\cite{herring}, they are known as the ``$Q$-approach" and 
the ``$\Pi$-approach". The equivalency of the two approaches is confirmed by 
the Onsager symmetry and the fundamental relationships between these approaches 
have been established in recent past\cite{Tsaou}.  In the ``$Q$-approach" 
a weak temperature gradient ${\bs \nabla}T$ is applied so that electrons and phonons 
move in directions opposite to each other. The flow of electrons causes the 
diffusion thermopower $S_d$. As a consequence of electron-phonon interaction
a finite fraction of momentum is transferred from phonons to electrons which drags
electrons in the opposite direction and phonon-drag contribution to the thermoelectric power 
comes into the picture. Several authors\cite{cantrell1, cantrell2, lyo2, smith}
have calculated phonon-drag thermopower in various electronic systems in this 
approach by solving coupled Boltzmann equations for both electrons and phonons. 
On the other hand in the ``$\Pi$-approach"
a very weak electric field ${\bf E}$ is applied to cause electrons drift. In this case, 
since no temperature gradient is applied (i.e. ${\bs \nabla}T=0$), phonons are 
in equilibrium. Interaction between electrons and phonons leads to 
transfer of momentum from electrons to phonons which produces a finite phonon heat
current. In this way phonon-drag contribution to the Peltier coefficient can be 
determined. Many authors\cite{gerin, lyo1, kuba, muli, fromhold} calculated 
phonon-drag thermopower of a 2DES in a perpendicular magnetic field using 
$\Pi$- approach.

Recently, 2DES with spin-orbit interaction\cite{zutic, winkler, cahay} (SOI) has 
become an emerging area of research due to its potential application for developing 
spintronic devices\cite{wolf, david, fabian}. Two types of SOI namely Rashba\cite{rashba}
and Dresselhaus\cite{dress} are present in low-dimensional semiconducting
structures. Rashba spin-orbit interaction (RSOI) occurs due to the inversion asymmetry 
of hetero-interface. An external gate voltage can tune\cite{nitta, mats} the strength 
of RSOI. Zero-field spin splitting\cite{luo, das, hu} is an important consequence 
of RSOI. On the other hand, SOI of Dresselhaus type occurs in crystals which have 
bulk inversion asymmetry. The RSOI has many important consequences on various 
properties of 2DES.
Electron-phonon interaction strength can be modified by RSOI\cite{cape, grima}. 
It causes an increase in polaron mass correction\cite{zhang}. Temperature dependence of 
phonon-limited mobility\cite{chen} and resistivity\cite{ghosh} get modified by RSOI.
The peak arising in longitudinal magnetoresistivity of a 2DES 
splits\cite{wang} into two due to RSOI. Very recently,
a thermoelectric probe\cite{firoz} has been used theoretically to calculate the 
strength of RSOI by analyzing the beating patterns obtained in the 
thermoelectric coefficients.

In this paper we calculate phonon-drag thermopower of a 2DES confined in 
a GaAs/AlGaAs heterostructures in presence of both perpendicular magnetic field 
and RSOI at very low temperatures. An oscillatory behaviour of phonon-drag thermopower 
with the applied magnetic field is found. It is found that at higher values of magnetic
field strong Rashba coupling is able to split the peaks appearing in phonon-drag thermopower.
This kind of splitting is considered as a direct effect of RSOI. The number of 
oscillations in the thermopower increases with electron density also. We study the behaviour of 
phonon-drag thermopower with temperature for various values of magnetic field, electron density and Rashba 
spin-orbit coupling constant. At very low temperature (BG) regime a power-law 
dependence of phonon-drag thermopower is observed. The exponents of this temperature 
dependence have been evaluated numerically for different parameters. 
It is established that the RSOI causes a strong suppression in the effective 
exponent of the phonon-drag thermopower.

This paper is organized as follows. In section II we present detailed theoretical 
calculations. In section III numerical results and discussions are given. 
We summarize our work in sec IV. Some calculations are shown in detail in 
the Appendices.

\section{Theory}

\subsection{Basic information of the physical system}

We consider a quasi-2DES confined at the interface of a GaAs/AlGaAs heterostructure.
Electrons are restricted to move in the $x$-$y$ plane due to a confining 
potential of triangular type in the growth direction (say, the $z$-direction).
We assume that only the lowest subband in the $z$-direction is occupied.
The total wave function can be written as $\Psi({x,y,z})=\Psi(x,y)\xi_0(z)$, 
where $\xi_0=\sqrt{b^3/2}ze^{-bz/2}$ is the Fang-Howard\cite{fang} variational 
wave function in the $z$-direction. The variational parameter is given by 
$b=(48\pi m^\ast e^2/\varepsilon_0 \kappa\hbar^2)^{1/3}\Big(n_d+11n_e/32\Big)^{1/3}$,
where $m^\ast$, $\kappa$, $\varepsilon_0$, $n_d$ and $n_e$ are effective mass of 
an electron in GaAs, dielectric constant of GaAs, permittivity
of free space, depletion charge density and density of electron, respectively.

The single particle Hamiltonian\cite{wang, lange} of a Rashba spin-orbit coupled 2DES
in the presence of a perpendicular magnetic field along $z$-direction can be written as

\begin{eqnarray}\label{hamil}
H=\frac{{\bf{P}}^2}{2m^\ast}+\frac{\alpha}{\hbar}\Big(\sigma_xP_y-\sigma_yP_x\Big)
+\frac{1}{2}g^\ast\mu_BB\sigma_z,
\end{eqnarray}
where ${\bf{P}}={\bf p}+e{\bf A}$ with ${\bf A}$ as the vector potential, $\alpha$ is 
the Rashba spin-orbit coupling constant, $\sigma_i$'s are the 
usual Pauli spin matrices, $g^\ast$ is the effective Lande g-factor and $\mu_B$ is 
Bohr magneton.
To solve Eq. (\ref{hamil}) we consider the Landau gauge
${\bf A}=(0,Bx,0)$.

The eigen spectrum is given by
\begin{eqnarray}\label{eigen_val}
\epsilon_n^{\lambda}=n\hbar\omega_c+\lambda\sqrt{\epsilon_0^2+2n\frac{\alpha^2}{l_0^2}},
\ \ \ \ n=1,2,\cdots
\end{eqnarray} 
where $\lambda=\pm$, $\omega_c=eB/m^\ast$ is the 
cyclotron frequency and $l_0=\sqrt{\hbar/eB}$ is the magnetic length.
For $n=0$ there is only one level with energy $\epsilon_0=(\hbar\omega_c-g^\ast\mu_BB)/2$.

The eigenfunctions corresponding to $\epsilon_n^\lambda$ are respectively given by
\begin{eqnarray}
\Psi_n^+({x,y})=\frac{e^{ik_yy}}{\sqrt{2\pi A_n}}\left(
\begin{array}{c}
D_n\phi_{n-1}\Big(\frac{x+x_0}{l_0}\Big)\\
\phi_n\Big(\frac{x+x_0}{l_0}\Big)\\
\end{array}\right)
\end{eqnarray}
and
\begin{eqnarray}
\Psi_n^-({x,y})=\frac{e^{ik_yy}}{\sqrt{2\pi A_n}}\left(
\begin{array}{c}
\phi_{n-1}\Big(\frac{x+x_0}{l_0}\Big)\\
-D_n\phi_n\Big(\frac{x+x_0}{l_0}\Big)\\
\end{array}\right).
\end{eqnarray}
Here $k_y$ is the $y$-component of electron wave vector ${\bf k}$,
$x_0=k_yl_0^2$ and
$\phi_n[(x+x_0)/l_0]= \sqrt{1/(2^nn!\sqrt{\pi})} e^{-(x+x_0)^2/(2l_0^2)}
H_n[(x+x_0)/l_0]$ 
is the harmonic oscillator wavefunction centered at $x=-x_0$.
The coefficients $D_n$ and $A_n$ are given by 
$D_n=\sqrt{2n}\alpha/l_0/(\epsilon_0+\sqrt{\epsilon_0^2+2n\alpha^2/l_0^2})$ and
$A_n=1+D_n^2$.                             
The eigenspinor corresponding to $n=0$ state is 
\begin{eqnarray}
\Psi_0^+({\bf r})=\frac{e^{ik_yy}}{\sqrt{2\pi}}\left(
\begin{array}{c}
0\\
\phi_0\Big(\frac{x+x_0}{l_0}\Big)\\
\end{array}\right).
\end{eqnarray}

The density of states (DOS) is given by 
$D(\epsilon)=(1/2\pi l_0^2)\sum_{n,\lambda}\delta(\epsilon-\epsilon_n^\lambda)$.
Considering Lorentzian broadening of the Landau levels 
the DOS can be written as 
\begin{eqnarray}\label{dos}
 D(\epsilon)=\frac{1}{2\pi^2l_0^2}\sum_{n,\lambda}
\frac{\Gamma_L}{(\epsilon-\epsilon_n^\lambda)^2+\Gamma_L^2},
\end{eqnarray}
where $\Gamma_L$ is the Lorentzian broadening parameter.

The chemical potential $\mu$ can be determined by the following
condition
\begin{eqnarray}
 n_e=\int_0^\infty d\epsilon D(\epsilon) f^0(\epsilon),
\end{eqnarray}
where $f^0(\epsilon)=1/(e^{\beta(\epsilon-\mu)}+1)$ with 
$\beta=1/(k_BT)$ is the usual Fermi-Dirac distribution function.

\subsection{Phonon-drag Thermopower}

To calculate phonon-drag thermopower we follow the ``$\Pi$-approach''
as described in Reference\cite{fromhold}. Since there is no temperature 
gradient (i.e. ${\bs \nabla}T=0$) the transport equations are 
quite simpler as 
${\bf J}=\overleftrightarrow{\sigma}{\bf E}$ and 
${\bf U}=\overleftrightarrow{\Pi}{\bf J}$, 
where ${\bf J}$, ${\bf U}$, $\overleftrightarrow{\sigma}$ 
and $\overleftrightarrow{\Pi}$ are electron current density,
phonon heat current density, conductivity tensor and 
Peltier coefficient tensor, respectively.
According to Kelvin, the thermopower and the Peltier coefficient are 
related thermodynamically as $T\overleftrightarrow{S}=\overleftrightarrow{\Pi}$.

The phonon heat current density is given by
\begin{eqnarray}\label{heat}
{\bf U}=\frac{1}{L^2}\sum_{{\bf q},s} \hbar\omega_{qs}{{\bf v}_{qs}}N_q^1,
\end{eqnarray}
where $L^2$ is the area of the sample, the index $s$ represents particular 
phonon mode, ${\bf v}_{qs}$ is the 
phonon velocity, $\omega_{qs}$ is the phonon frequency and finally the shift in 
phonon distribution is $N_q^1=N_q-N_q^0$.

The steady-state Boltzmann equation can be used to find $N_q^1$ as 
\begin{eqnarray}\label{Boltz}
\Big(\frac{\partial N_q}{\partial t}\Big)_{\rm ep}+
\Big(\frac{\partial N_q}{\partial t}\Big)_{\rm coll}=0.
\end{eqnarray}
Here, the 1$^{st}$ term represents the rate of change in phonon distribution due to 
electron-phonon interaction and 2$^{nd}$ term arises from various 
scattering processes such as phonon-phonon scattering, surface 
roughness scattering etc.
In the relaxation time approximation the 2$^{nd}$ term of Eq. (\ref{Boltz}) 
can be written as 
\begin{eqnarray}\label{relx}
\Big(\frac{\partial N_q}{\partial t}\Big)_{\rm coll}=
-\frac{N_q-N_q^0}{\tau_p}=-\frac{N_q^1}{\tau_p},
\end{eqnarray}
where $\tau_p$ is the phonon relaxation time.
Substitution of Eq. (\ref{relx}) into Eq. (\ref{Boltz}) yields 
\begin{eqnarray}\label{eq_lin}
N_q^1=\tau_p\Big(\frac{\partial N_q}{\partial t}\Big)_{\rm ep}.
\end{eqnarray}

Now the rate of change in phonon distribution function due to 
electron-phonon interaction is given by 

\begin{eqnarray}\label{relx1}
\Big(\frac{\partial N_q}{\partial t}\Big)_{\rm ep}&=&
\sum_{\nu,\nu^\prime}
\Big[P_{\nu^\prime\nu}^{\rm em}
f_{\nu^\prime}(\epsilon_{{\nu^\prime}})\{1-f_\nu(\epsilon_\nu)\}\nonumber\\
&-& P_{\nu\nu^\prime}^{\rm ab}
f_\nu(\epsilon_{\nu})\{1-f_{\nu^\prime}(\epsilon_{\nu\prime})\}\Big],
\end{eqnarray}
where $\nu \equiv (n,k_y,\lambda)$ represents the set of quantum numbers, 
$f_i(\epsilon_i)$'s are the electron distribution functions,
$P_{\nu\nu^\prime}^{\rm ab}(P_{\nu\nu^\prime}^{\rm em})$
is the probability which is responsible for making
transition of an electron from the initial state $\nu$ to the 
final state $\nu^\prime$ with the absorption(emission) of a 
phonon.

According to the Fermi's golden rule we have 

\begin{eqnarray}
P_{\nu\nu^\prime}^{\rm ab(\rm em)} & = &
\frac{2\pi}{\hbar}\vert M_{\nu\nu^\prime}({\bf q})\vert^2
\Big(N_q^0+\frac{1}{2}\mp\frac{1}{2}\Big) \nonumber \\
& \times & \delta\Big(\epsilon_{\nu^\prime}-\epsilon_{\nu}\mp\hbar\omega_q\Big), 
\end{eqnarray}
where $\vert M_{\nu\nu^\prime}({\bf q})\vert^2$ is square of the matrix 
element responsible for the electron-phonon interaction and 
$N_q^0=1/(e^{\beta\hbar\omega_q}-1)$ is the equilibrium Bose distribution 
function. Finally $+$ and $-$ signs in the 
parentheses represent emission and absorption, respectively. A detailed
description of $\vert M_{\nu\nu^\prime}({\bf q})\vert^2$ is given in Appendix A.

The applied electric field is vanishingly small 
so that one can linearize Eq. (\ref{eq_lin}) about the equilibrium value.
To do this we write the electron distribution function as 
$f_{\nu(\nu^\prime)}=f_{\nu(\nu^\prime)}^0+f_{\nu(\nu^\prime)}^1$. 
So Eq. (\ref{eq_lin}) can be written as 
\begin{eqnarray}\label{relx2}
N_q^1=\frac{\tau_p}{k_BT}\sum_{\nu\nu^\prime}
\Big(\frac{f_\nu^1}{f_{\nu}^\prime}
-\frac{f_{\nu^\prime}^1}{f_{\nu^\prime}^\prime}\Big)
W_{\nu^\prime\nu},
\end{eqnarray}
where $f_{\nu(\nu^\prime)}^\prime
=\frac{\partial f_{\nu(\nu^\prime)}^0}{\partial\epsilon_{\nu(\nu^\prime)}}$
and $W_{\nu^\prime\nu}=f_\nu^0(1-f_{\nu^\prime}^0)P_{\nu\nu^\prime}^{\rm ab}$.

Since the applied electric field is very small, it can be treated as 
a perturbative term in the Hamiltonian. Using the 1$^{st}$ order perturbation 
theory the energy eigenvalue of a state $\nu$ in the presence of ${\bf E}$ 
can be modified as $\epsilon_{\nu(\nu^\prime)}\simeq\epsilon_{\nu(\nu^\prime)}^0
+eE\langle x\rangle_{\nu(\nu^\prime)}$, where $\epsilon_{\nu(\nu^\prime)}^0$ is the 
unperturbed energy spectrum given by Eq. (\ref{eigen_val}). 
The expectation values of $x$ are given by 
$\langle x\rangle_{\nu(\nu^\prime)}=
\langle\psi_{\nu(\nu^\prime)}({\bf r})\vert x\vert\psi_{\nu(\nu^\prime)}({\bf r})\rangle
=-l_0^2k_y(k_y^\prime)$.

Therefore, Eq. (\ref{relx2}) can be written as 
\begin{eqnarray}
N_q^1=\frac{\tau_peE}{k_BT}\sum_{\nu\nu^\prime}
\Big\{\langle x\rangle_{\nu^\prime}-\langle x\rangle_{\nu}\Big\}W_{\nu^\prime\nu}.
\end{eqnarray}
 
The components of the 
heat current density can be written as 
\begin{eqnarray}\label{pnheatx}
U_x=\frac{e\tau_pE}{k_BTL^2}\sum_{\nu,\nu^\prime,{\bf q},s}
\hbar\omega_{qs}\Big\{\langle x\rangle_{\nu\prime}-\langle x\rangle_\nu\Big\}
W_{\nu^\prime\nu}v_{qx}
\end{eqnarray}
and
\begin{eqnarray}\label{pnheaty}
U_y=\frac{e\tau_pE}{k_BTL^2}\sum_{\nu,\nu^\prime,{\bf q},s}
\hbar\omega_{qs}\Big\{\langle x\rangle_{\nu^\prime}-\langle x\rangle_\nu\Big\}
W_{\nu^\prime\nu}v_{qy}.
\end{eqnarray}
Now the conservation of momentum in $y$-direction forces us to write 
$\langle x\rangle_{\nu^\prime}-\langle x\rangle_\nu=-q_yl_0^2$.
Also we have $v_{qx(y)}=v_sq_{x(y)}/q$. 
With these substitution  Eqs. (\ref{pnheatx}) and (\ref{pnheaty}) become
\begin{eqnarray}\label{ht_crtx}
U_x=-\frac{e\tau_pl_0^2E}{k_BTL^2}\sum_{\nu,\nu^\prime,{\bf q},s}
\hbar\omega_{qs}v_s W_{\nu^\prime\nu}\frac{q_xq_y}{q}
\end{eqnarray}
and
\begin{eqnarray}\label{ht_crty}
U_y=-\frac{e\tau_pl_0^2E}{k_BTL^2}\sum_{\nu,\nu^\prime,{\bf q},s}
\hbar\omega_{qs}v_s W_{\nu^\prime\nu}\frac{q_y^2}{q}.
\end{eqnarray}

Now $\sum_{\bf q}\rightarrow (1/(2\pi)^3)\int dq_{\parallel}q_\parallel
d\phi dq_z$, where $q_\parallel=\sqrt{q_x^2+q_y^2}$.
If we evaluate all the summations involved in the above equations 
one can readily obtain $U_x=0$ and this is quite obvious because
the application of an electric field along $x$ direction causes a drift 
current in $y$ direction in presence of a magnetic field in $z$ direction.
For a particular phonon mode $s$,
$U_y$ is given by (detailed calculations are given in Appendix B)
\begin{eqnarray}\label{heat_current}
U_y&=&-\frac{e\Lambda_pv_s\Gamma_L^2E}{8\pi^4k_BT}\sum_{n,\lambda}
\int dq_{\parallel} dq_z q_{\parallel}^3
\hbar\omega_q N_q^0(N_q^0+1)\nonumber\\
&\times&\frac{\vert C_{{\bf q}s}\vert^2F_{nn}^\lambda(q_\parallel) I_z(q_z)}
{\{(\epsilon_F-\epsilon_n^\lambda)^2+\Gamma_L^2\}
\{(\epsilon_F-\epsilon_n^\lambda+\hbar\omega_{qs})^2+\Gamma_L^2\}},\nonumber\\
\end{eqnarray}
where $\vert C_{{\bf q}s}\vert^2$, $F_{nn}(q_\parallel)$ and $I_z(q_z)$ are the 
square of the matrix element for various mechanisms of electron-phonon interaction, 
the in-plane and out-of-plane form factors, respectively. The phonon mean free path 
is given by $\Lambda_p=v_s\tau_p$ which is nearly $0.3$ mm in the present case.
In deriving Eq. (\ref{heat_current})
we have considered only the intra-Landau level (i.e. $n=n^\prime$) and 
intra-branch (i.e. $\lambda=\lambda^\prime$) scattering because
at very low temperature inter-Landau level\cite{lyo1} and inter-branch
contributions to the thermopower are negligibly small.

Now the phonon-drag thermopower\cite{kuba, muli} can be written as 
\begin{eqnarray}\label{phn_drag}
 S_{xx}=S_{yy}=\frac{1}{T}\frac{U_y}{E}\rho_{xy},
\end{eqnarray}
where the Hall resistivity in presence of RSOI is given by \cite{firoz1} 
$\rho_{xy}\simeq(B/n_ee)(1+k_\alpha^2/k_F^2)$ with 
$k_\alpha=m^\ast\alpha/\hbar^2$ and $k_F=\sqrt{2\pi n_e}$.

\section{Numerical results and discussions}

In this section we present all the numerical results obtained by solving 
Eq. (\ref{phn_drag}) numerically. For the numerical calculations we adopt the values 
of material parameters appropriate for GaAs as $m^\ast=0.067m_0$ with $m_0$ is the mass 
of a free electron, $n_0=10^{15}$ m$^{-2}$, 
$\alpha_0=10^{-11}$ eV-m, $n_d=5\times10^{14}$ m$^{-2}$, $\kappa=12.91$,
$v_{sl}=5.12\times10^3$ ms$^{-1}$, 
$v_{st}=3.04\times10^3$ ms$^{-1}$, $D=12$ eV, $h_{14}=1.2\times10^9$ V-m$^{-1}$, 
$\rho_m=5.31\times 10^3$ Kg-m$^{-3}$.
The Landau level broadening parameter $\Gamma_L$ depends on various parameters 
like magnetic field, temperature etc. For simplicity, we have taken here a constant value of
$\Gamma_L$ as $\Gamma_L=1.2$ meV.

\begin{figure}[t]
\begin{center}\leavevmode
\includegraphics[width=150mm]{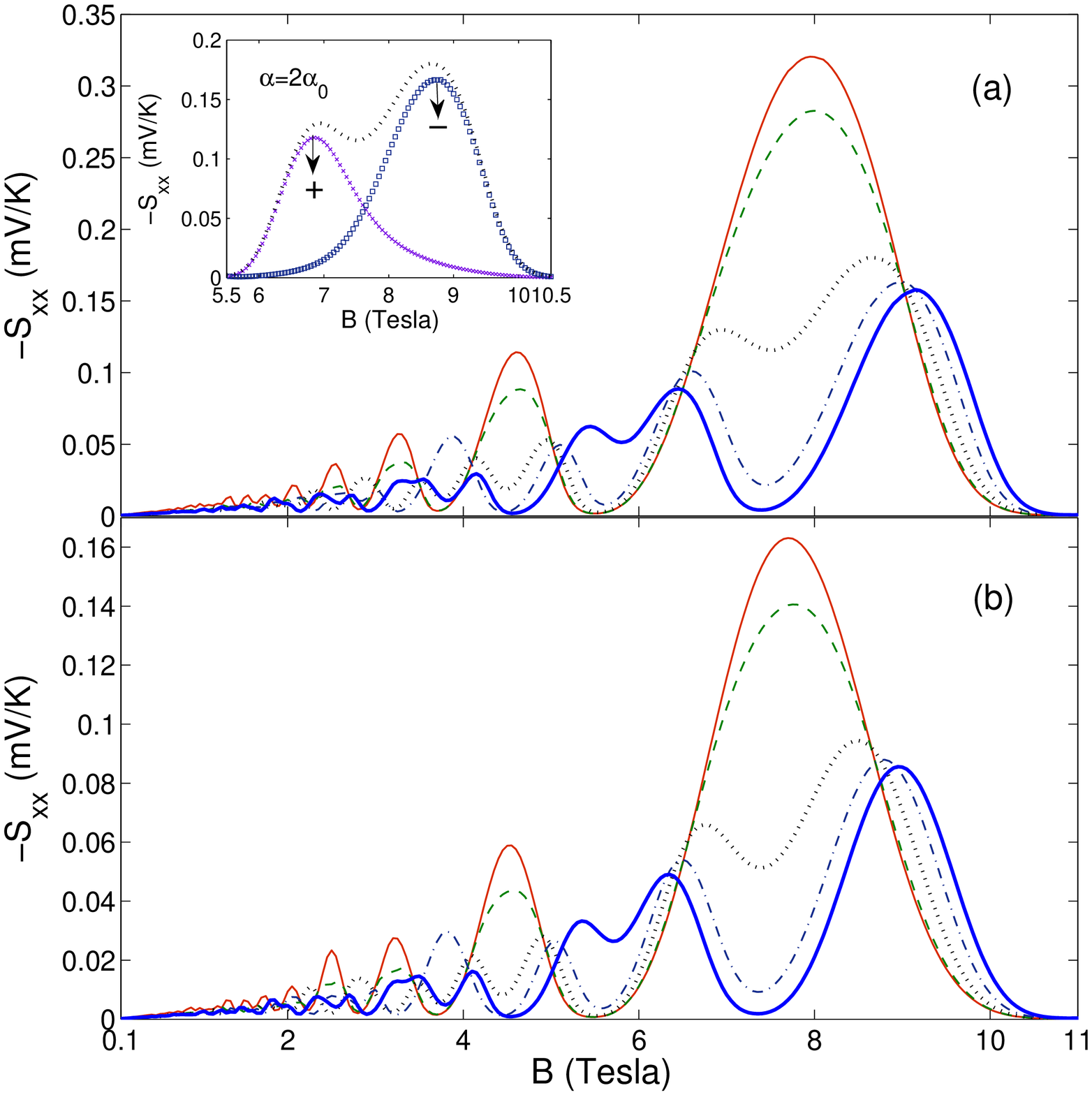}
\caption{(Color online) Plots of the phonon-drag thermopower vs 
magnetic field at a fixed density $n_e=5n_0$ and a fixed temperature 
$T=2$ K. The upper and lower panels are drawn for DP and PE scattering, 
respectively. Here, solid, dashed, dotted, dash-dotted and thick-solid 
lines represent $\alpha=0, \alpha_0, 2\alpha_0, 3\alpha_0$ and $4\alpha_0$, 
respectively. Contributions to $S_{xx}$ from different branch of Rashba spectrum
for $\alpha=2\alpha_0$ are shown in the inset.}
\label{Fig3}
\end{center}
\end{figure}

\begin{figure}[]
\begin{center}\leavevmode
\includegraphics[width=150mm]{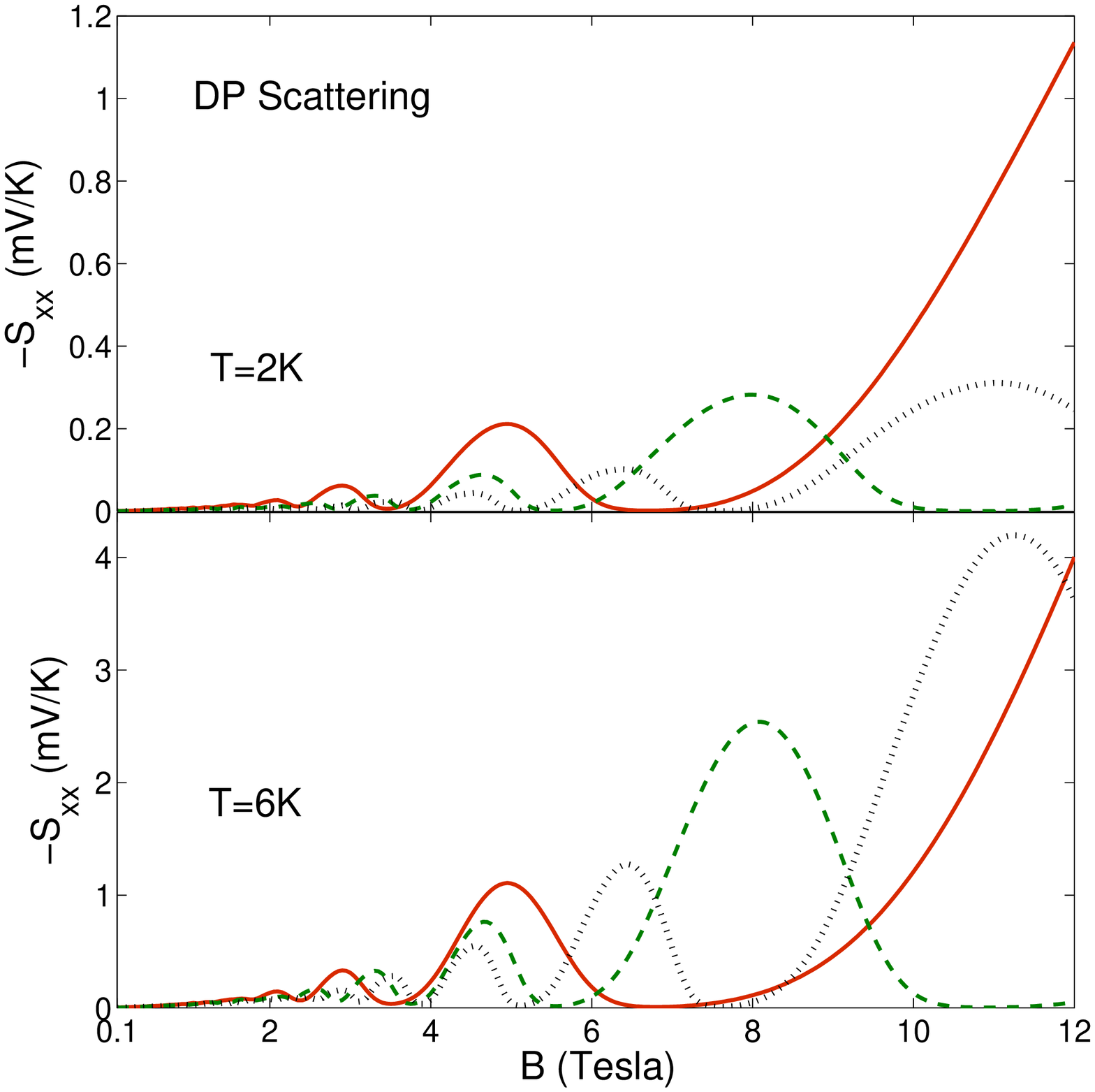}
\caption{(Color online) Plots of the phonon-drag thermopower vs  magnetic field
due to DP scattering at a fixed $\alpha=\alpha_0$. Upper and lower panels are, 
respectively, for $T=2$ K and $T=6$ K. Here, solid, dashed and dotted lines correspond 
to $n_e=3n_0, 5n_0$ and $7n_0$, respectively.}
\label{Fig3}
\end{center}
\end{figure}

\begin{figure}[]
\begin{center}\leavevmode
\includegraphics[width=150mm]{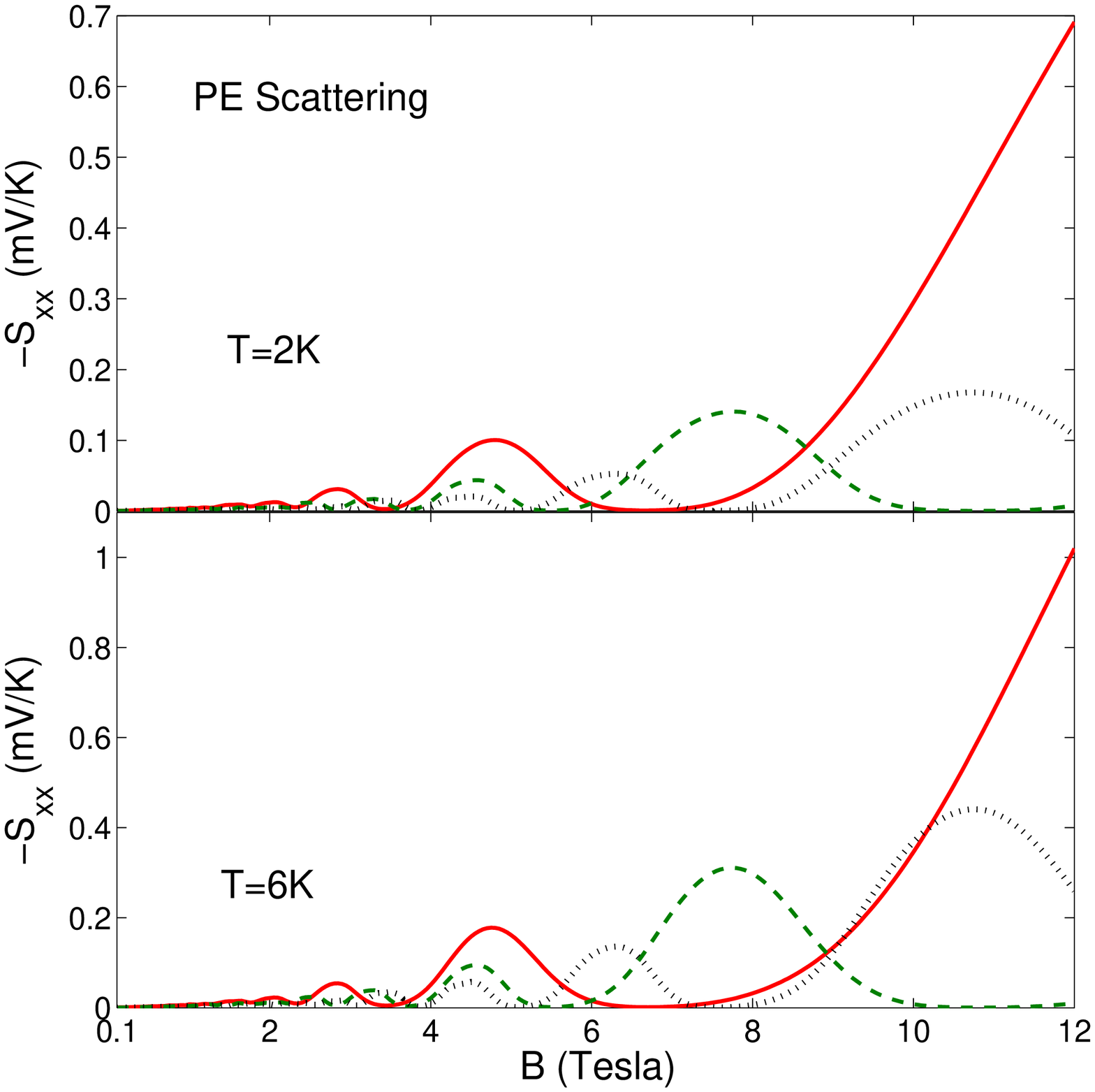}
\caption{(Color online) Plots of the phonon-drag thermopower vs  magnetic field
due to PE scattering at a fixed $\alpha=\alpha_0$. Upper and lower panels are,
respectively, for $T=2$ K and $T=6$ K. Here, solid, dashed and dotted lines correspond 
to $n_e=3n_0, 5n_0$ and $7n_0$, respectively.}
\label{Fig3}
\end{center}
\end{figure}

\begin{figure}[]
\begin{center}\leavevmode
\includegraphics[width=110mm]{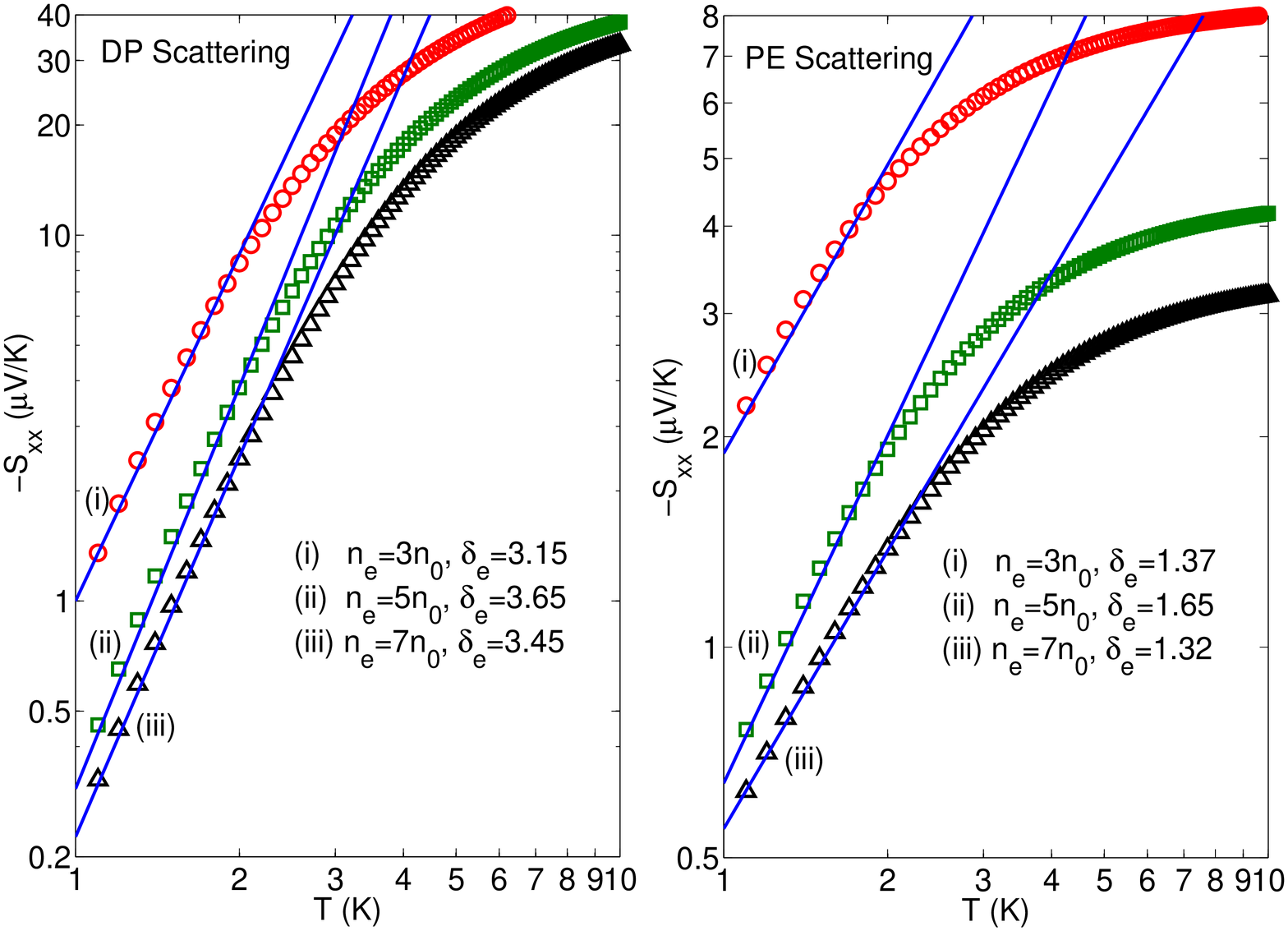}
\caption{(Color online) Plots of the phonon-drag thermopower vs  temperature
due to both DP and PE scattering at a fixed $\alpha=\alpha_0$ and $B=1$ T.}
\label{Fig4}
\end{center}
\end{figure}

\begin{figure}[]
\begin{center}\leavevmode
\includegraphics[width=110mm]{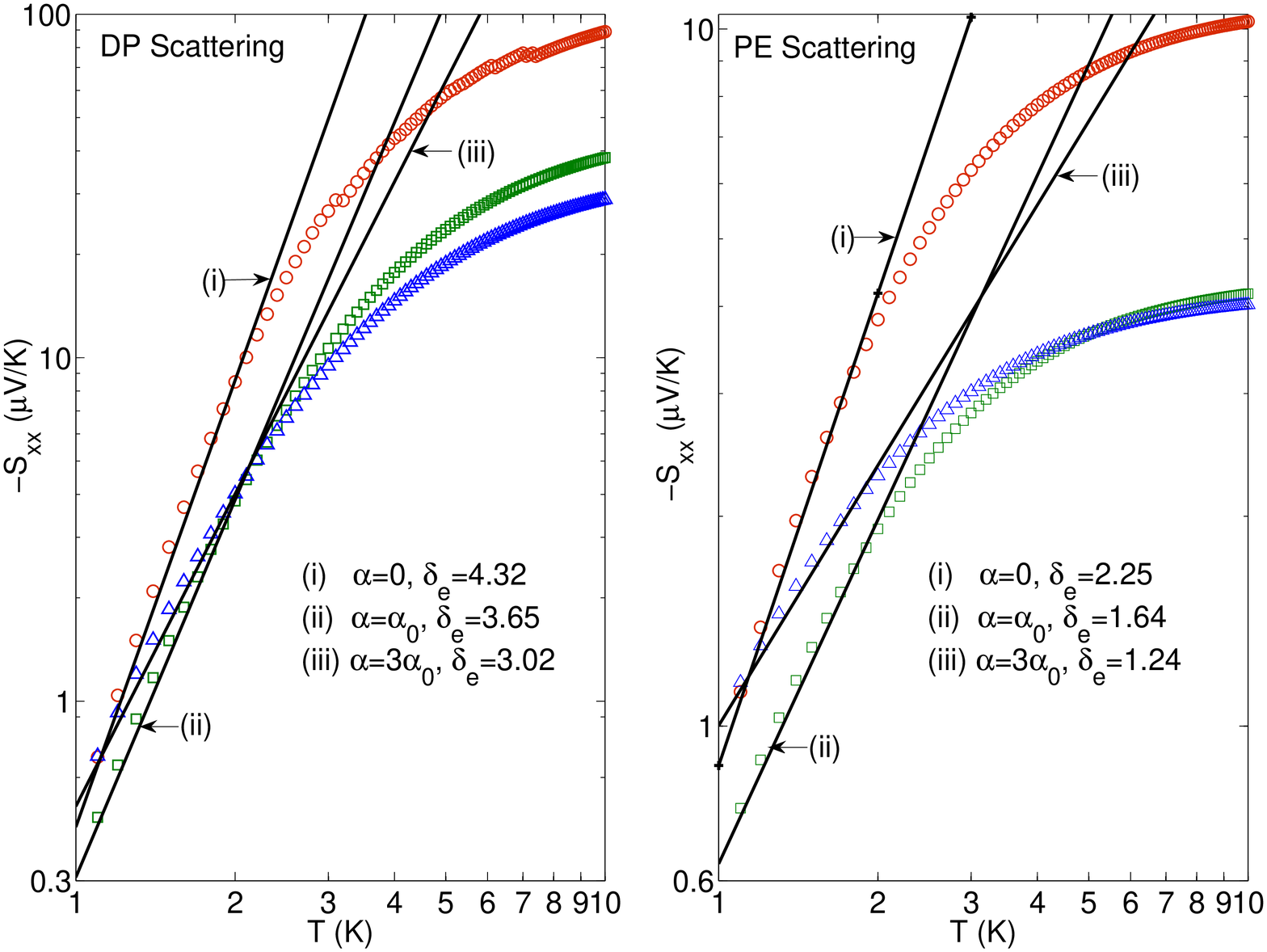}
\caption{(Color online) Plots of the phonon-drag thermopower vs  temperature
due to both DP and PE scattering at a fixed density $n_e=5n_0$ and $B=1$ T.}
\label{Fig4}
\end{center}
\end{figure}

\begin{figure}[]
\begin{center}\leavevmode
\includegraphics[width=110mm]{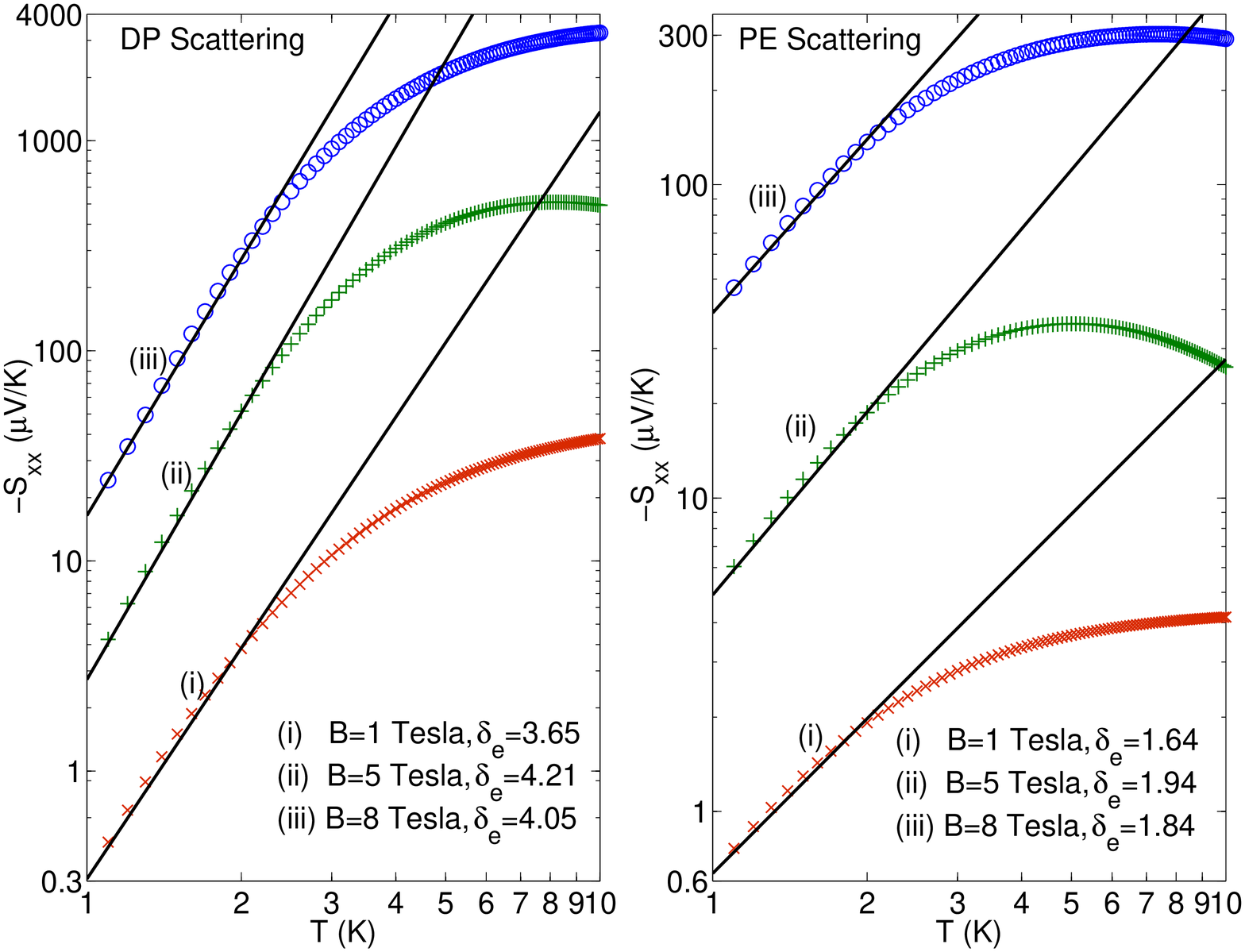}
\caption{(Color online) Plots of the phonon-drag thermopower vs  temperature
due to both DP and PE scattering at a fixed $\alpha=\alpha_0$ and $n_e=5n_0$.}
\label{Fig4}
\end{center}
\end{figure}

In Fig. 1 we present the dependence of the phonon-drag thermopower on the 
applied magnetic field at a fixed density $n_e=5n_0$ and a fixed temperature  
$T=2$ K. Two panels $(a)$ and $(b)$ of Fig. 1 represent DP and PE scattering, 
respectively. Different curves are plotted for different values of $\alpha$.
It can be seen that the amplitude of $S_{xx}$ decreases gradually with the 
increase of $\alpha$. One interesting fact is that at higher values of
magnetic field the peaks in $S_{xx}$ corresponding to the $\alpha=0$
split into two as $\alpha$ increases. The two split peaks are well separated
from each other at higher values of $\alpha$ (say, $\alpha=4\alpha_0$).
To explain the origin of this splitting of peaks physically, a zoomed 
portion of Fig. 1$(a)$ for $\alpha=2\alpha_0$ in the range of the the magnetic 
field $B=(5.5-10.5)$ T is shown in the inset of Fig. 1$(a)$. 
This splitting is an important consequence of the contributions coming
from the two branches of the Rashba spin-split energy spectrum. 
Contributions to the phonon-drag thermopower from ``$+$'' branch and 
``$-$'' branch are shown in the inset. The splitting occurs as a 
result of the finite phase difference between two contributions. Also the  
contributions are not equal in magnitude. Comparing the two panels 
$(a)$ and $(b)$ we conclude that 
the magnitude of phonon-drag thermopower due to DP scattering is greater 
than that due to PE scattering.

In Fig. 2 we plot $S_{xx}$ due to DP scattering as a function of $B$ at a fixed 
$\alpha=\alpha_0$ for different densities: $n_e = 3n_0$, $5n_0$ and $7n_0$. Here, 
we consider two different temperatures $T=2$ K and $T=6$ K. With the increase of 
density number of oscillations increases. The magnitude of $S_{xx}$ is higher at 
higher value of temperature. In Fig. 3 we plot the same as in Fig. 2 for PE 
scattering. The rate of increment in $S_{xx}$ due to PE scattering  with temperature 
is slower than that due to DP scattering.

The temperature dependence of phonon-drag thermopower 
for various densities is shown in Fig. 4. We have plotted 
$S_{xx}$ as a function of temperature $T$ for a fixed value of magnetic field
$B=1$ T and the Rashba spin-orbit coupling constant $\alpha=\alpha_0$. \
We consider both DP and PE scattering mechanisms of electron-phonon interaction 
separately. In the range of temperature up to $T\simeq3$ K the phonon-drag
thermopower shows a power-law dependence $S_{xx}\sim T^{\delta_e}$. We extract 
the effective exponent of the temperature dependence from the log-log plot of 
$S_{xx}$ versus $T$ as shown in Fig. 4. We find due to DP scattering 
$\delta_e=3.15$, $3.65$ and $3.45$ for $n_e=3n_0$,$5n_0$ and $7n_0$,
respectively. Due to PE scattering it is found that $\delta_e=1.37$, $1.65$ 
and $1.32$ for $n_e=3n_0$, $5n_0$ and $7n_0$, respectively. The exponent $\delta_e$
depends on density. In Fig. 5 we describe the temperature dependence of $S_{xx}$ for 
different values of $\alpha$ at fixed magnetic field $B=1$ T and 
fixed density $n_e=5n_0$. We have found the exponent $\delta_e=4.32, 3.65$ and $3.02$
for $\alpha=0, \alpha_0$ and $3\alpha_0$, respectively due to DP scattering.
For PE scattering we find  $\delta_e=2.25, 1.64$ and $1.24$
for $\alpha=0, \alpha_0$ and $3\alpha_0$, respectively.
Analyzing Fig. 5 we conclude that the presence of strong RSOI 
suppresses the exponents of the temperature dependence of $S_{xx}$ significantly.
Similar results were found recently\cite{biswas} in the temperature dependence of the
phonon-drag thermopower of a Rashba spin-orbit coupled 2DES with zero magnetic field.
In Fig. 6 we plot $S_{xx}$ as a function of $T$ for various values of $B$. In this 
case we fix $n_e=5n_0$ and $\alpha=\alpha_0$. Exponents of the temperature dependence of 
$S_{xx}$ have been calculated in this case also. For DP scattering we find
$\delta_e=3.65, 4.21$ and $4.05$ for $B=1, 5$ and $8$ T, respectively. 
We have $\delta= 1.64, 1.94$ and $1.84$ for $B=1, 5$ and $8$ T, respectively,
due to PE scattering. In our previous study\cite{biswas} with $B=0$, $n_e=5n_0$
and $\alpha=\alpha_0$ we found $\delta_e=3.294$ for DP scattering and 
$\delta_e=1.520$ for PE scattering. So one can conclude that a finite amount of magnetic 
field is able to introduce a significant enhancement in the effective exponent of the 
temperature dependence of phonon-drag thermopower.

\section{Summary}
In summary, we have studied phonon-drag contribution to the thermoelectric
power of a two-dimensional electron system confined in a GaAs/AlGaAs heterostructure
in presence of both Rashba spin-orbit interaction and perpendicular magnetic field.
We have considered both deformation potential and piezoelectric scattering mechanisms 
of electron-phonon interaction. Interaction between electrons with two-dimensional 
wave vector and phonons of three dimensional wave vector has been taken into 
consideration. Dependence of phonon-drag thermopower on an external magnetic field 
and temperature has been discussed thoroughly. At higher values of magnetic field 
splitting of peaks in phonon-drag thermopower is found at strong values of 
Rashba spin-orbit coupling constant. This splitting of peak is a direct 
effect of the Rasha spin-orbit interaction. We have also found a power-law dependence
of phonon-drag thermopower on temperature.  
It is found that the exponent strongly depends on elctron density, magnetic field and the spin-orbit
coupling constant.

\appendix
\section{}

The Hamiltonian describing electron-phonon interaction can be written as 
\begin{eqnarray}
 H_{\rm ep}=\sum_{{\bf q}, s}(C_{{\bf q}s}e^{i{\bf q}\cdot{\bf r}}a_{{\bf q}s}+
 C_{{\bf q}s}^\dagger e^{-i{\bf q}\cdot {\bf r}}a_{{\bf q}s}^\dagger),
\end{eqnarray}
where $a_{{\bf q}s}(a_{{\bf q}s}^\dagger)$ is the phonon annihilation (creation)
operator and $C_{\bf{q}s}$ is the matrix element responsible for electron-phonon
interaction in a particular phonon mode $({\bf q},s)$.

The square of the matrix element of $H_{\rm ep}$ is given by
\begin{eqnarray}\label{app_matrix}
\vert M_{\nu,\nu^\prime}({\bf q})\vert^2&=&
\vert\langle\psi_\nu({\bf r})\vert H_{\rm ep}
\vert\psi_{\nu^\prime}({\bf r})\rangle\vert^2\nonumber\\
&=&\vert C_{{\bf q}s}\vert^2F_{nn^\prime}^\lambda(q_\parallel)
I_z(q_z).
\end{eqnarray}
Here, the electron-phonon matrix elements ($C_{{\bf q}s}$) are 
different for various scattering mechanisms. For DP and PE scattering 
the square of the matrix elements ($\vert C_{{\bf q}s}\vert^2$) are 
respectively given by 
$\vert C_{{\bf q}l}^{\rm DP}\vert^2=D^2\hbar q/(2\rho_mv_{sl})$ and 
$\vert C_{{\bf q},{l(t)}}^{\rm PE}\vert^2=
(eh_{14})^2\hbar A_{l{(t)}}({q_\parallel},q_z)/(4\rho_mv_{s{l(t)}}q)$, 
where $D$ is deformation potential constant, $h_{14}$ is the relevant PE 
coupling tensor component, $\rho_m$ is the mass density and $v_{sl(t)}$ is 
the longitudinal(transverse) component of sound velocity. 
Finally the anisotropy factors in the longitudinal and transverse directions 
are given by $A_l=9q_\parallel^4q_z^2/2q^6$ 
and $A_t=(8q_\parallel^2q_z^4+q_\parallel^6)/4q^6$, respectively.

In Eq. (\ref{app_matrix}),
$F_{nn^\prime}^\lambda(q_\parallel)=\vert\langle\psi_\nu({\bf r})\vert e^{i{q_xx+q_yy}}
\vert\psi_{\nu^\prime}({\bf r})\rangle\vert^2$ is the in-plane form factor.
For upper and lower branch we have, respectively, 

\begin{eqnarray}
 F_{nn^\prime}^+(q_\parallel)=B_n^{n^\prime}(\zeta)
\Big[\sqrt{\frac{n}{n^\prime}}D_nD_{n^\prime}L_{n^\prime-1}^{n-n^\prime}(\zeta)
+L_{n^\prime}^{n-n^\prime}(\zeta)\Big]^2
\end{eqnarray}
and 
\begin{eqnarray}
F_{nn^\prime}^-(q_\parallel)=B_n^{n^\prime}(\zeta)
\Big[\sqrt{\frac{n}{n^\prime}}L_{n^\prime-1}^{n-n^\prime}(\zeta)
+D_nD_{n^\prime}L_{n^\prime}^{n-n^\prime}(\zeta)\Big]^2,
\end{eqnarray}
where $B_n^{n^\prime}=(n^\prime!/n!)\zeta^{n-n^\prime}e^{-\zeta}
\delta_{k_y^\prime, k_y+q_y}$
with $\zeta=q_\parallel^2l_0^2/2$.
 
The out-of-plane form factor is given by
$I_z(q_z)=\vert\langle \xi_0(z)\vert e^{iq_zz}\vert\xi_0(z)\rangle\vert^2
=b^6/(q_z^2+b^2)^3$.

\section{}
The summations involved in Eqs. (\ref{ht_crtx}) and (\ref{ht_crty}) 
can be written as 
\begin{eqnarray}\label{app_sum}
\sum_{\nu,\nu^\prime,{\bf q}, s}\longrightarrow
\sum_{n,n^\prime,k_y,k_y^\prime,\lambda,\lambda^\prime,{\bf q},s}.
\end{eqnarray}
The summation over $k_y^\prime$ can be easily done using the
Kronecker delta symbol $\delta_{k_y^\prime, k_y+q_y}$ in
$\vert M_{\nu,\nu^\prime}\vert^2$ arising from the momentum conservation
along $y$ direction. Again we have $\sum_{k_y}=L^2/2\pi l_0^2$.
Then Eq. (\ref{app_sum}) is simplified to 
\begin{eqnarray}
\sum_{\nu,\nu^\prime,{\bf q}, s}\overset{k_y^\prime=k_y+q_y}\longrightarrow
\frac{L^2}{2\pi l_0^2}\sum_{n,n^\prime,\lambda,\lambda^\prime,{\bf q},s}.
\end{eqnarray}
Now one can convert the summation over ${\bf q}$ as 
$\sum_{\bf q}\rightarrow (1/(2\pi)^3)\int dq_{\parallel}q_\parallel
d\phi dq_z$ with $q_\parallel=\sqrt{q_x^2+q_y^2}$ and $\phi=\tan^{-1}(q_y/q_x)$.
Equation (\ref{ht_crty}) can be written as 
\begin{eqnarray}\label{heat_cur}
U_y&=&-\frac{e\tau_pE}{8\pi^2k_BT}\sum_{n,n^\prime,\lambda,\lambda^\prime s}
v_s^2\int dq_\parallel dq_z q_\parallel^3\vert M_{\nu,\nu^\prime}(q)\vert^2
N_q^0\nonumber\\
&\times& f^0(\epsilon_n^\lambda)\{1-f^0(\epsilon_n^\lambda+\hbar\omega_{qs})\}
\delta(\epsilon_{n^\prime}^{\lambda^\prime}-\epsilon_{n}^{\lambda}-\hbar\omega_{qs}).\nonumber\\
\end{eqnarray}
Using
$\delta(\epsilon_{n^\prime}^{\lambda^\prime}-\epsilon_{n}^{\lambda}-\hbar\omega_{qs})
=\int d\epsilon\delta(\epsilon-\epsilon_{n}^{\lambda})
\delta(\epsilon-\epsilon_{n^\prime}^{\lambda^\prime}+\hbar\omega_{qs})$,
Eq. (\ref{heat_cur}) becomes 
\begin{eqnarray}\label{heat_app}
U_y&=&-\frac{e\tau_pE}{8\pi^2k_BT}\sum_{n,n^\prime,\lambda,\lambda^\prime s}
v_s^2\int dq_\parallel dq_z q_\parallel^3\vert M_{\nu,\nu^\prime}(q)\vert^2
N_q^0G_{nn^\prime}^{\lambda\lambda^\prime}\nonumber\\
\end{eqnarray}
with 
$G_{nn^\prime}^{\lambda\lambda^\prime}=\int d\epsilon\delta(\epsilon-\epsilon_{n}^{\lambda})
\delta(\epsilon-\epsilon_{n^\prime}^{\lambda^\prime}+\hbar\omega_{qs})
f^0(\epsilon)\{1-f^0(\epsilon+\hbar\omega_{qs})\}.$

Now in the presence of disorder broadening of Landau level occurs. 
Assuming Lorentzian broadening of width $\Gamma_L$ the delta-function 
in the expression of 
$G_{nn^\prime}^{\lambda\lambda^\prime}$ can be written as 
$\delta(\epsilon-\epsilon_{n}^{\lambda})=
(1/\pi)\Gamma_L/((\epsilon-\epsilon_{n}^{\lambda})^2+\Gamma_L^2)$
as described in Eq. (\ref{dos}) also. At very low temperature where 
phonon energy $\hbar\omega_{qs}$ is comparable with the thermal energy 
$k_BT$ and $\hbar\omega_{qs}, k_BT<<\epsilon_F$
we can make the following approximation 
$f^0(\epsilon)\{1-f^0(\epsilon+\hbar\omega_{qs})\}
\simeq\hbar\omega_{qs}(N_q^0+1)\delta(\epsilon-\epsilon_F)$.
Also in the BG regime the intra-level and intra-branch Landau 
level scatterings dominate over inter-level scattering. One  
can write $n=n^\prime$ and $\lambda=\lambda^\prime$. Now 
it is straightforward to arrive at Eq. (\ref{heat_current}) from 
Eq. (\ref{heat_app}) by taking all the approximations into consideration.

\end{document}